\documentclass[journal]{IEEEtran}

\usepackage{graphicx}
\usepackage{siunitx}
\usepackage{balance}
\begin{document}

\title{Nanoseconds Timing System Based on IEEE 1588\\
FPGA Implementation}

\author{D.~Pedretti, M.~Bellato, R.~Isocrate, A.~Bergnoli, R.~Brugnera, D.~Corti, F.~Dal~Corso, G.~Galet, A.~Garfagnini, A.~Giaz, I.~Lippi, F.~Marini, G.~Andronico, V.~Antonelli, M.~Baldoncini, E.~Bernieri, A.~Brigatti,  A.~Budano, M.~Buscemi, S.~Bussino, R.~Caruso, D.~Chiesa,
C.~Clementi, X.~F.~Ding, S.~Dusini, A.~Fabbri, R.~Ford, A.~Formozov, M.~Giammarchi, M.~Grassi, A.~Insolia, P.~Lombardi, F.~Mantovani, S.~M.~Mari, C.~Martellini, A.~Martini, E.~Meroni, L.~Miramonti, S.~Monforte, P.~Montini, M.~Montuschi, M.~Nastasi, F.~Ortica, A.~Paoloni, E.~Previtali, G.~Ranucci, A.~C.~Re, B.~Ricci, A.~Romani, G.~Salamanna, F.~H.~Sawy, G.~Settanta, M.~Sisti, C.~Sirignano, L.~Stanco, V.~Strati, G.~Verde\\
\vspace{0.5cm}
\textit{On~Behalf~of~the~JUNO~Collaboration}

\thanks{D. Pedretti, is with the Department
of Information Engineering, Padova University and INFN Legnaro National Laboratories, Italy.}% <-this % stops a space

\thanks{M. Bellato, R. Brugnera, D. Corti, F. Dal Corso, S. Dusini, G. Galet, A. Garfagnini, A. Giaz, R. Isocrate, I. Lippi, F. Marini, F. H. Sawy, C. Sirignano, L. Stanco are with the Department of Physics and Astronomy, Padova University and INFN Padova, Italy.}

\thanks{A. Bergnoli is with DWave.it, Padova, Italy.}

\thanks{G. Andronico, S. Monforte, G. Verde are with INFN Catania, Italy.}

\thanks{V. Antonelli, A. Brigatti, A. Formozov, M. Giammarchi, P. Lombardi, E. Meroni, L. Miramonti, G. Ranucci, A. C. Re are with the Department of Physics and Astronomy, Milano University and INFN Milano, Italy.}

\thanks{M. Baldoncini, F. Mantovani, M. Montuschi, B. Ricci are with the Department of Physics and Earth Science, Ferrara University and INFN Ferrara, Italy.}

\thanks{E. Bernieri, A. Budano, S. Bussino, A. Fabbri, S. M. Mari, C. Martellini, P. Montini, G. Salamanna, G. Settanta, are with the Department of Physics and Mathematics, Roma Tre University and INFN Roma Tre, Italy.}

\thanks{M. Buscemi, R. Caruso, A. Insolia are with the Department of Physics and Astronomy, Catania University and INFN Catania, Italy.}

\thanks{D. Chiesa, M. Nastasi, E. Previtali, M.~Sisti are with the Department of Physics, Milano Bicocca University and INFN Milano Bicocca, Italy.}

\thanks{C. Clementi, F. Ortica, A. Romani are with the Department of Chemistry, Biology and Biotechnology, Perugia University and INFN Perugia, Italy.}

\thanks{A. Martini, A. Paoloni are with INFN Frascati National Laboratories, Frascati, Italy.}

\thanks{V. Strati is with the Department of Physics and Earth Science, Ferrara University and INFN Legnaro National Laboratories, Italy.}

\thanks{M. Grassi is with INFN Milano, Italy and APC Laboratory IN2P3, Paris, France.}
\thanks{R. Ford is with INFN Milano, Italy and SNOLAB, Ontario, Canada.}
\thanks{X. F. Ding is with INFN Milano and Gran Sasso Science Institute, L'Aquila, Italy.}
}

\IEEEpubid{0000--0000/00\$00.00˜\copyright˜2018 IEEE } 
\markboth{IEEE Transaction on Nuclear Science, June~2018}%
{Shell \MakeLowercase{\textit{et al.}}: Nanoseconds Timing System Based on IEEE 1588 FPGA Implementation}


\maketitle

\begin{abstract}
Clock synchronization procedures are mandatory in most physical experiments where event fragments are readout by spatially dislocated sensors and must be glued together to reconstruct key parameters (e.g. energy, interaction vertex etc.) of the process under investigation. These distributed data readout topologies rely on an accurate time
information available at the frontend, where raw data are acquired and tagged with a precise timestamp prior to data buffering and central data collecting. This makes the network complexity and latency, between frontend and backend electronics, negligible within upper bounds imposed by the frontend data buffer capability. The proposed research work describes an FPGA implementation of IEEE 1588 Precision Time Protocol (PTP) that exploits the CERN Timing, Trigger and Control (TTC) system as a multicast messaging physical and data link layer. The hardware implementation extends the clock synchronization to the nanoseconds range, overcoming the typical accuracy limitations inferred by computers Ethernet based Local Area Network (LAN). 
Establishing a reliable communication between master and timing receiver nodes is essential in a message-based synchronization system.
In the backend electronics, the serial data streams synchronization with the global clock domain is guaranteed by an hardware-based finite state machine that scans the bit period using a variable delay chain and finds the optimal sampling point.
The validity of the proposed timing system has been proved in point-to-point data links as well as in star topology configurations over standard CAT-5e cables. The results achieved together with weaknesses and possible improvements are hereby detailed.
\end{abstract}


\IEEEpeerreviewmaketitle

\IEEEpubidadjcol

\section{Introduction}
\IEEEPARstart{T}{he} context of the proposed research work is the Jiangmen Underground Neutrino Observatory (JUNO) \cite{JUNO1} \cite{JUNO2}. The timing system is an essential part of this experiment, indeed, to precisely determine the energy and the interaction vertex of incident neutrinos, the charge information coming from the 18000 photomultipliers (PMTs) surrounding the central detector must be associated with a precise time information.
The JUNO readout architecture foresees the frontend electronics, hereby represented by the Global Control Unit (GCU) card, to be installed underwater, close to the PMTs \cite{GCU}.
The heart of the frontend electronics is an FPGA that handles data digitization, data buffering, data readout, slow control and monitoring, trigger generation and synchronization with the backend electronics. Trigger requests generated by different channels are asynchronous and independent events that must be attached with a timestamp in order to be correctly processed by the central trigger system. The central trigger system collects trigger requests coming from all the readout channels and generates trigger validations. The trigger validation is essentially a data readout request delivered to all the GCUs with the goal of collecting data fragments in a time window centered around a \textit{center$\_$time} parameter specified in the request message. The \textit{center$\_$time} must be univocally interpreted by the central trigger system and by all the GCUs. This evidences how the distributed nature of the data readout demands for a synchronization system whose primary task is to handle an accurate time distribution from the central timing and trigger system to all the frontend cards. The trigger synchronization accuracy demanded is $\mathit{\pm}$ 16 ns.

\subsection{Different Approaches to Synchronization}

There are essentially two different approaches to synchronization widely used in physics experiments: event-based synchronization and time-based synchronization \cite{comparison}. 
Both are valid solutions that emphasize different concepts of synchronization and both potentially lead to an accuracy in the sub-ns range, upon precise delay and asymmetry measurement and compensation.

\subsubsection{Event-based Synchronization}
The main task of an event-based timing system is to deliver reliable, fixed and low latency control messages to all the nodes reached by the synchronization network. An example of event-based synchronization is the Micro-Research Finland (MRF) timing system \cite{MRF}. In this topology, the event generator is the only holder of the global time information and it converts the scheduled timing events in optical signals delivered through a deterministic network to an array of event receivers. Each receiver converts the received event codes to synchronized digital output pulses.
The automatic delay measurement and compensation is a key feature to achieve a sub-ns resolution. 

\subsubsection{Time-based Synchronization}
The primary task of a time-based timing system is to handle an accurate time distribution and clock synchronization. The precision time protocol, defined in the IEEE 1588-2008 standard, is an example of time-based synchronization since it relies on an accurate copy of the global time held in thousands of timing receiver nodes \cite{1588 PTP}. The accuracy of a PTP software implementation over a standard Ethernet LAN rarely extends in the sub-\SI{}{\micro s} range.

White Rabbit (WR) timing system is an example of time-based synchronization that exploits the IEEE 1588-2008 standard and extends the timing resolution to the sub-ns range featuring a 1000 base-LX Synchronous Ethernet over single-mode optical fiber and implementing a phase tracking system based on a Digital Dual-Mixer Time Difference (DDMTD) phase detection \cite{White Rabbit}. 

\section{The JUNO Synchronization Scheme}
\begin{figure}[!t]
\centering
\includegraphics*[width=90mm]{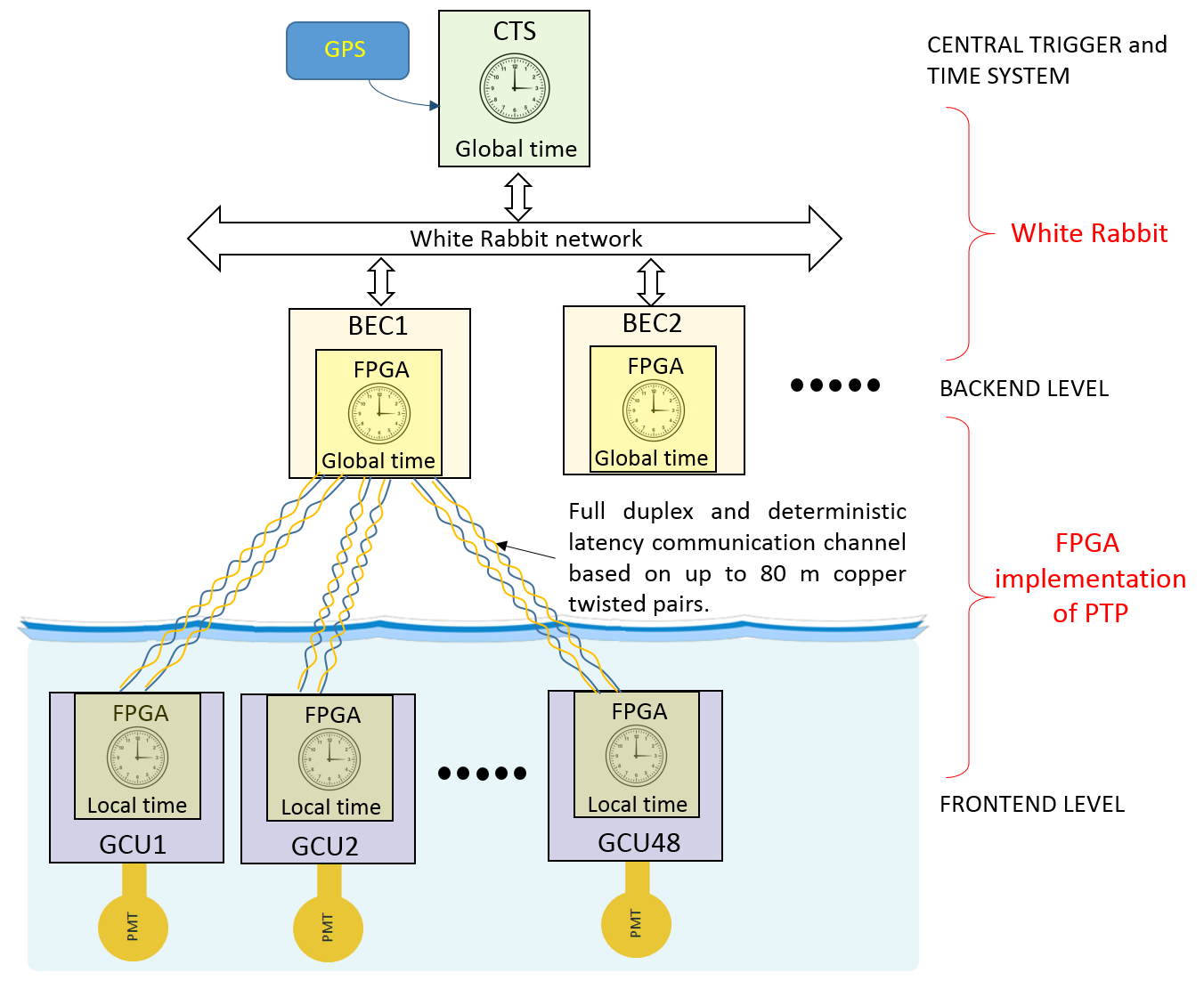}
\caption{JUNO synchronization scheme overview.}
\label{pedre1}
\end{figure}

The choice of the timing system to be used depends on which services are demanded to the synchronization system and on the data readout architecture.
In JUNO, the main task of the timing system is to generate an accurate copy of the global time at the backend and frontend levels of the readout architecture as shown in Figure \ref{pedre1}.
The Central Trigger and Timing System (CTS) holds the global time to be distributed to all the timing receivers. The backend electronics cards (BECs) are integrated into the White Rabbit network that may provide a sub-ns synchronization with the CTS.
The global clock signal must be distributed from the BECs to the frontend cards that count the time locally. 
Every local count will experience an offset with respect to the global time counter since the start of the counting is not synchronized among GCUs. This offset must be measured and corrected. 

The WR network ends at the BEC level and cannot extend to the frontend electronics since the potting of underwater electronics imposes tight constraints on the number of communication channels between the BEC and each GCU as well as on the communication medium. The JUNO collaboration did not envisage the use of optical fibers underwater 
and the adoption of WR on copper cables would nullify the benefits of the phase tracking procedure foreseen by WR, degrading its resolution.
The communication medium between BEC and GCU, on which is based this research work, is restricted to a couple of 80 m long twisted pairs in a CAT-5e cable.

The first clock alignment proposal was a synchronous reset pulse to be sent from the CTS to all the GCUs. This solution has been discarded since the delay calibration and compensation over asymmetric communication media, like the CAT-5e cable, needs special hardware to be accomplished.
Moreover, the synchronous reset solution is not selective and it would preclude the possibility to execute the clock alignment procedure runtime. If for any reason, one channel loses the synchronization, the operator should exclude that channel from the data readout until the next run of the experiment.

The offset correction mechanism between backend and frontend electronics, object of this paper, is based on three pillars:
\begin{itemize}
\item an hardware implementation of the IEEE 1588-2008 standard;
\item clock syntonization based on the clock data recovery (CDR) strategy;
\item the implementation of a full duplex and deterministic latency communication link layer between BEC and GCU over two copper twisted pairs.
\end{itemize} 
The paper details the advantages and disadvantages of the proposed solution supported by the experimental results achieved.

\section{PTP theory and synchronization performances} 

\subsection{Clock Offset Correction Mechanism}

\begin{figure}[!t]
\centering
\includegraphics*[width=78mm]{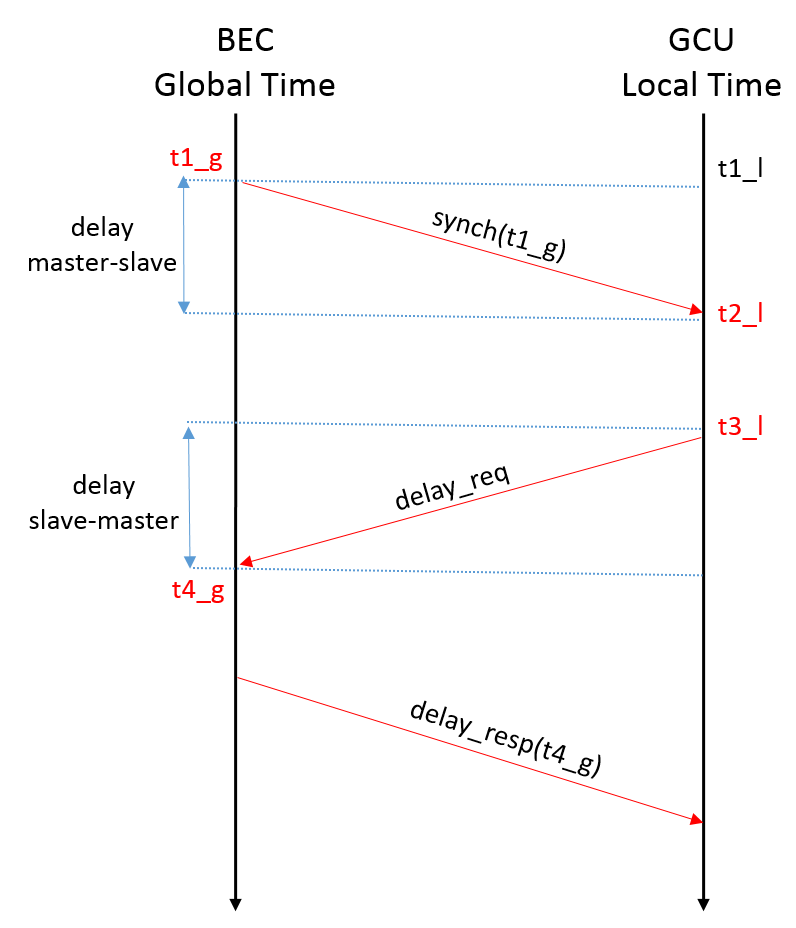}
\caption{IEEE 1588 PTP offset measurement \textsuperscript{\textcopyright}2008 IEEE.}
\label{pedre2}
\end{figure}

IEEE 1588-2008 standard defines a protocol for precise clock synchronization applicable to systems that implement a multicast communication model between the master and the timing nodes. 
In JUNO, assuming a messaging exchange capability, the idea is to exploit the delay request-response mechanism measurement defined in the IEEE 1588-2008 standard to compensate for the offset error between backend and frontend electronics. Figure \ref{pedre2} shows the protocol implemented. The follow-up message is not strictly necessary and is not used in the proposed solution.
$\mathit{t_{1\_g}} - \mathit{t_{1\_l}}$ is the clock offset to be measured and compensated. The offset measurement procedure is accomplished in eight steps: 
\begin{enumerate}
\item The master records the current timestamp $\mathit{t_{1\_g}}$ and sends to the slave a \textit{synch} messages containing the timestamp $\mathit{t_{1\_g}}$.
\item The slave records the reception time $\mathit{t_{2\_l}}$. The slave computes: $\mathit{t_{1\_g}} - \mathit{t_{2\_l}} = \mathit{offset} - \mathit{delay_{ms}}$ where $\mathit{delay_{ms}}$ is the transmission delay from master to slave.
\item The slave sends a delay request message, without payload, to the master and records the transmission time $\mathit{t_{3\_l}}$.
\item The master records the reception time $\mathit{t_{4\_g}}$.
\item The master sends back a delay message containing $\mathit{t_{4\_g}}$ value.
\item The slave, upon receipt of the \textit{delay\_resp} message computes: $\mathit{t_{4\_g}} - \mathit{t_{3\_l}} = \mathit{offset} + \mathit{delay_{sm}}$.
\item Now, with the assumption: $\mathit{delay_{ms}} = \mathit{delay_{sm}}$, the offset can be computed using (\ref{eq1}):
\begin{equation}
\mathit{offset} = \frac{(\mathit{t_{1\_g}} - \mathit{t_{2\_l}})+(\mathit{t_{4\_g}} - \mathit{t_{3\_l}})}{2}
\label{eq1}
\end{equation}
\item the slave corrects its clock accordingly.
\end{enumerate}

The master individually addresses the offset correction procedure to each slave. The procedure is periodical, thus ensuring clock alignment during the run of the experiment and offering the possibility to check the synchronization status of all the GCUs. Frequent offset corrections are indicative of problems and the corresponding GCU should be brought offline for diagnosis and firmware maintenance. 
 
\subsection{PTP Performances and Limiting Factors}
The offset correction mechanism highlights the sources of error that potentially bound the clock alignment accuracy:
\begin{itemize}
\item The protocol does not specify the clock frequency; lower-frequencies lead to poorer time resolutions.
\item Timestamping is a time critical operation. Hardware-assisted timestamping is required to achieve time synchronization in the ns range.
\item The synchronization over standard Ethernet LAN rarely goes beyond the \si{\micro}{s} of accuracy due to packet latency in the Ethernet network that is traffic dependent. The best performances of PTP over Ethernet are usually achieved with the Deterministic Ethernet, a communication technology that uses time scheduling to ensure a bounded and low latency transmission of the critical scheduled messages. 
The proposed timing system and the results described in this paper rely on a deterministic latency for each of the 18000 channels; PTP accounts for differences in these latencies.  
\item Like all message-based synchronization protocols, PTP time accuracy is degraded by asymmetry. Asymmetry usually originates from the physical medium and from the implementation of the data link layer. The assumption $\mathit{delay_{ms}} = \mathit{delay_{sm}}$ is not true in presence of asymmetry. Specifically, the time offset error is 1/2 of the asymmetry. 
\end{itemize}

\subsection{Advantages of a Digital Implementation}
\begin{figure}[!t]
\centering
\includegraphics*[width=90mm]{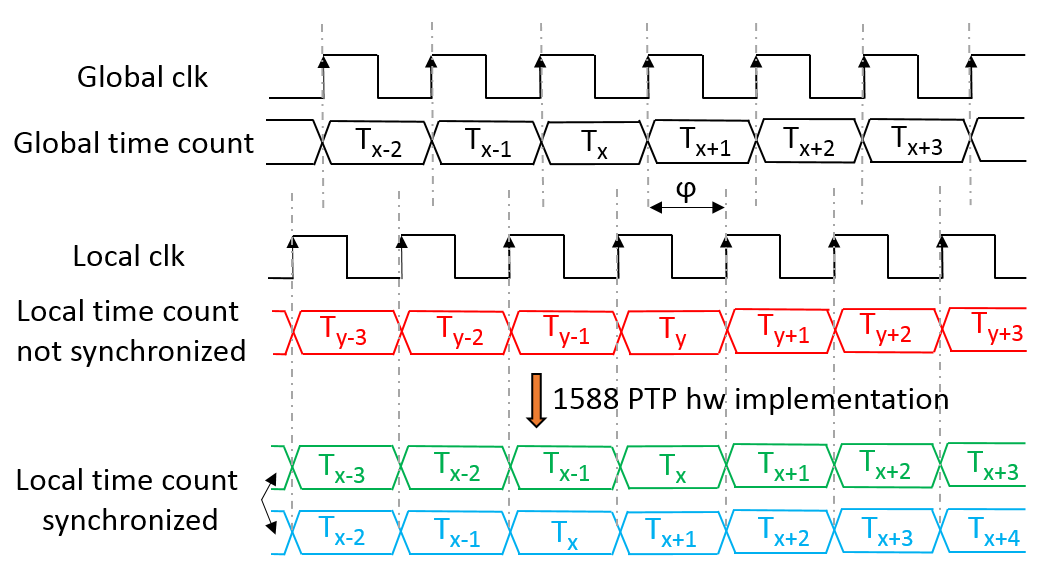}
\caption{Offset correction temporal diagram.}
\label{pedre3}
\end{figure}

Hardware-assisted implementations of PTP over Ethernet exist and prove that performing in hardware specific tasks leads to tight time synchronization \cite{phyter}. If the assumption of a perfect symmetry holds, the theoretical resolution of a fully digital implementation of PTP is $\mathit{\pm}$ one clock period as shown by the temporal diagram of Figure \ref{pedre3}.
The time is implemented in the form of a digital counter that counts the periods of the clock signal. The local time, prior to the clock alignment procedure, differs from the global time by a random offset $\mathit{T_y} - \mathit{T_x}$, determined by the power up sequence and by the time to lock of the phase-locked loop (PLL). The offset correction mechanism may end up in a configuration in which the local time lags (green) or leads (blue) the global time. Both are correct and acceptable solutions determined by the phase difference $\mathit{\varphi}$ between the global clock edge and the local clock edge; this phase difference $\mathit{\varphi}$ is mainly determined by the transmission latency. The PTP by construction cannot resolve $\mathit{\varphi}$, whose measurement is usually done via phase tracking systems (e.g. the DDMTD phase detection in WR). In the proposed timing system $\mathit{\varphi}$ is unknown but in principle, without variations of the cable length, it is invariant with a standard deviation imposed by the jitter.

\begin{figure}[!t]
\centering
\includegraphics*[width=89mm]{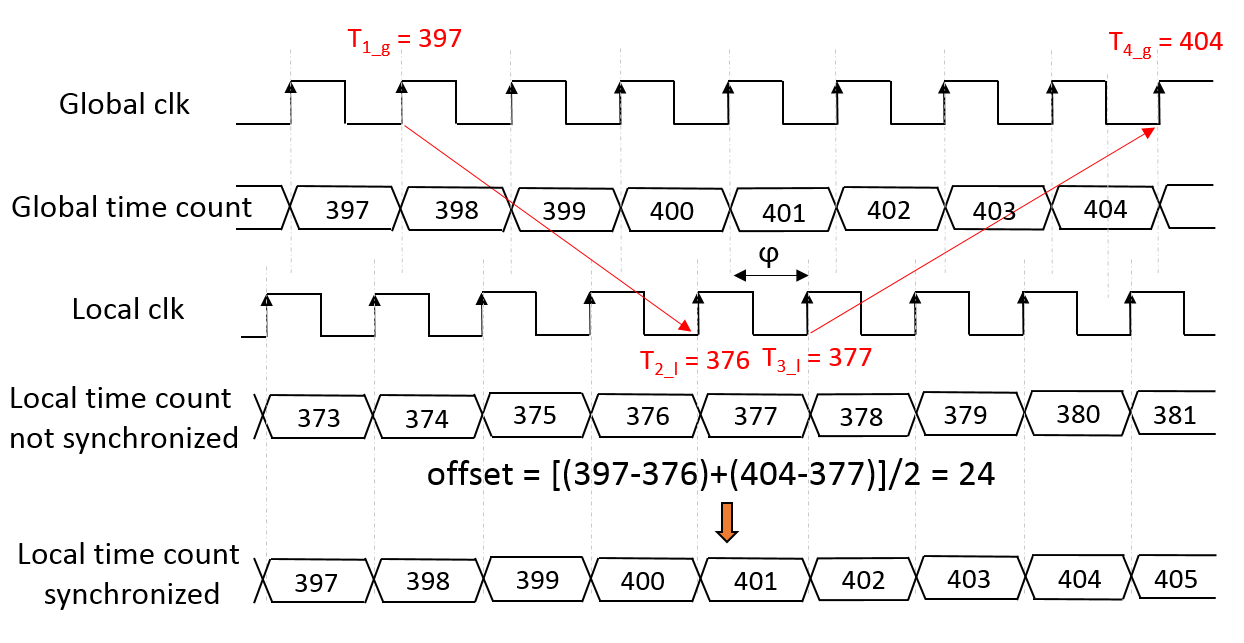}
\caption{Positive offset correction example.}
\label{pedre4}
\end{figure}

\begin{figure}[!t]
\centering
\includegraphics*[width=89mm]{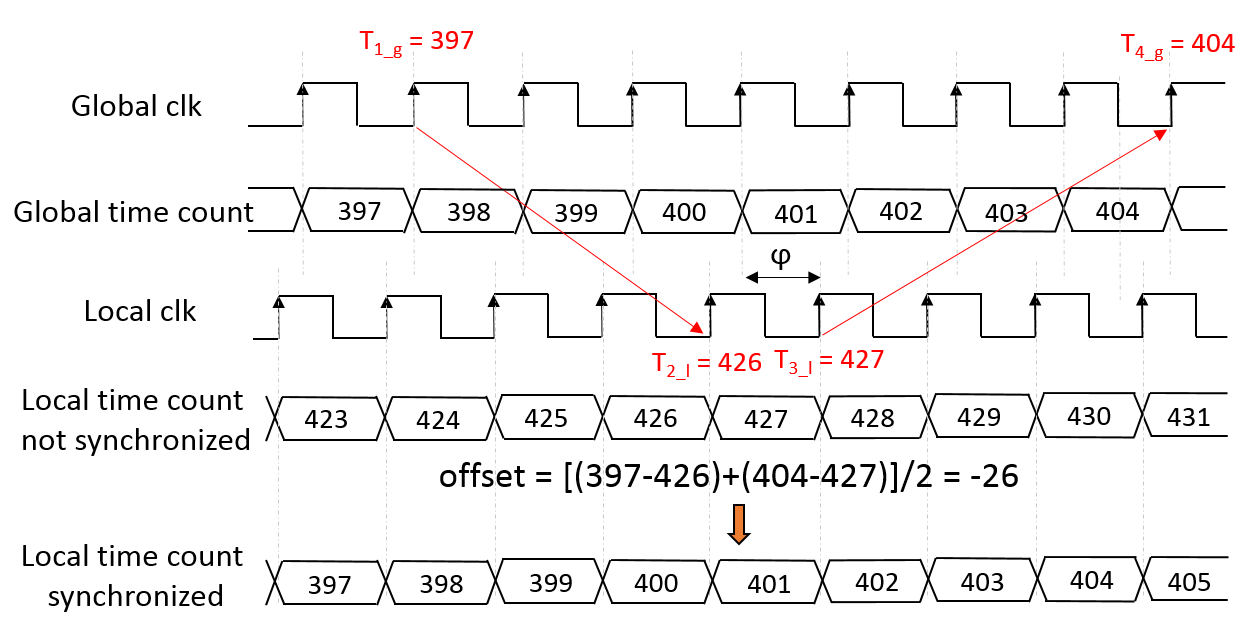}
\caption{Negative offset correction example.}
\label{pedre5}
\end{figure}

Two numerical examples of the offset correction are given in Figure \ref{pedre4} and Figure \ref{pedre5}. The red arrows represent the \textit{synch} and \textit{delay\_req} messages. The timestamps are registered on the clock rising edges that coincide with the messages transmission and/or reception. From these examples, we can deduce that with a 250 MHz global clock frequency the expected synchronization accuracy will be $\mathit{\pm}$ 4 ns.

\begin{figure}[!t]
\centering
\includegraphics*[width=89mm]{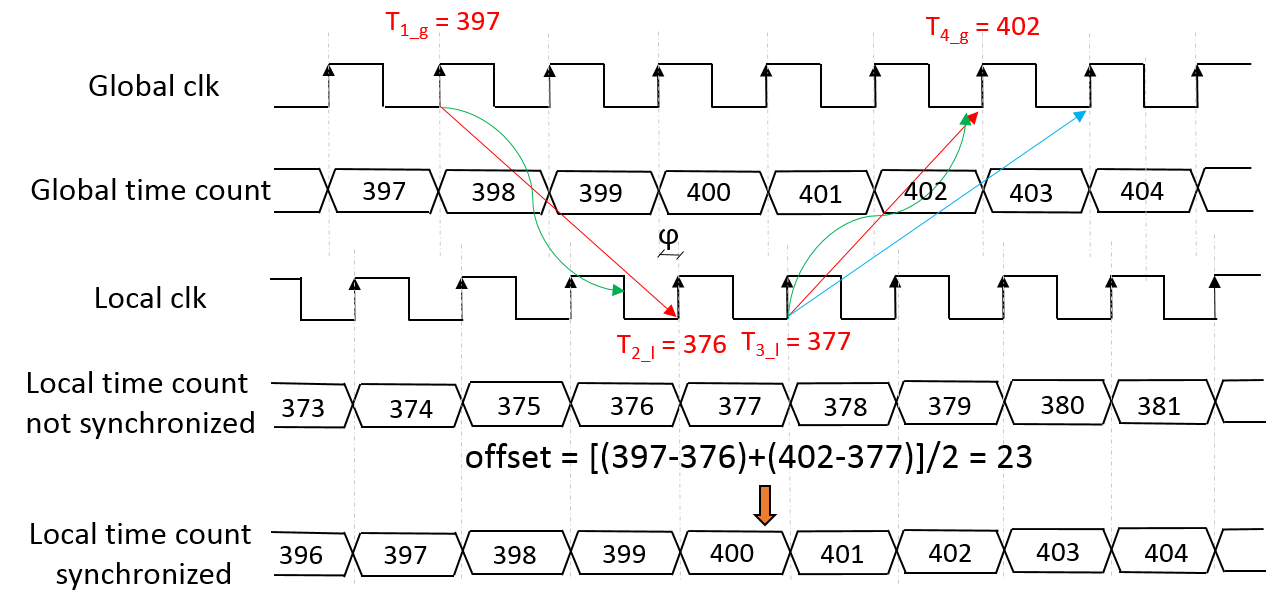}
\caption{The impact of data synchronization on the time accuracy.}
\label{pedre6}
\end{figure}

Figure \ref{pedre6} shows one more example with a different $\mathit{\varphi}$ that leads to a final configuration with the local count slightly lagging the global time. The picture also highlights the serial data synchronization issue and its impact on the time accuracy. As explained later in the paper, in the backend electronics, the input data stream have to be synchronized with the global clock domain. Indeed, if the data transition, indicated by the green arrowhead, is too close to the clock rising edge, an ambiguity in the received timestamp may arise due to timing violations. $\mathit{t_{4\_g}}$ might be 403, blue arrow, and the computed offset would be 23.5. Afterward, the division by two, implemented with a 1-bit right shift operation, rounds down the offset to 23. Any asymmetry contribution less than 2 clock periods does not affect the synchronization correctness.

\section{Timing system implementation}

\subsection{TTC as Physical and Data Link Layer}

\begin{figure}[!t]
\centering
\includegraphics*[width=91mm]{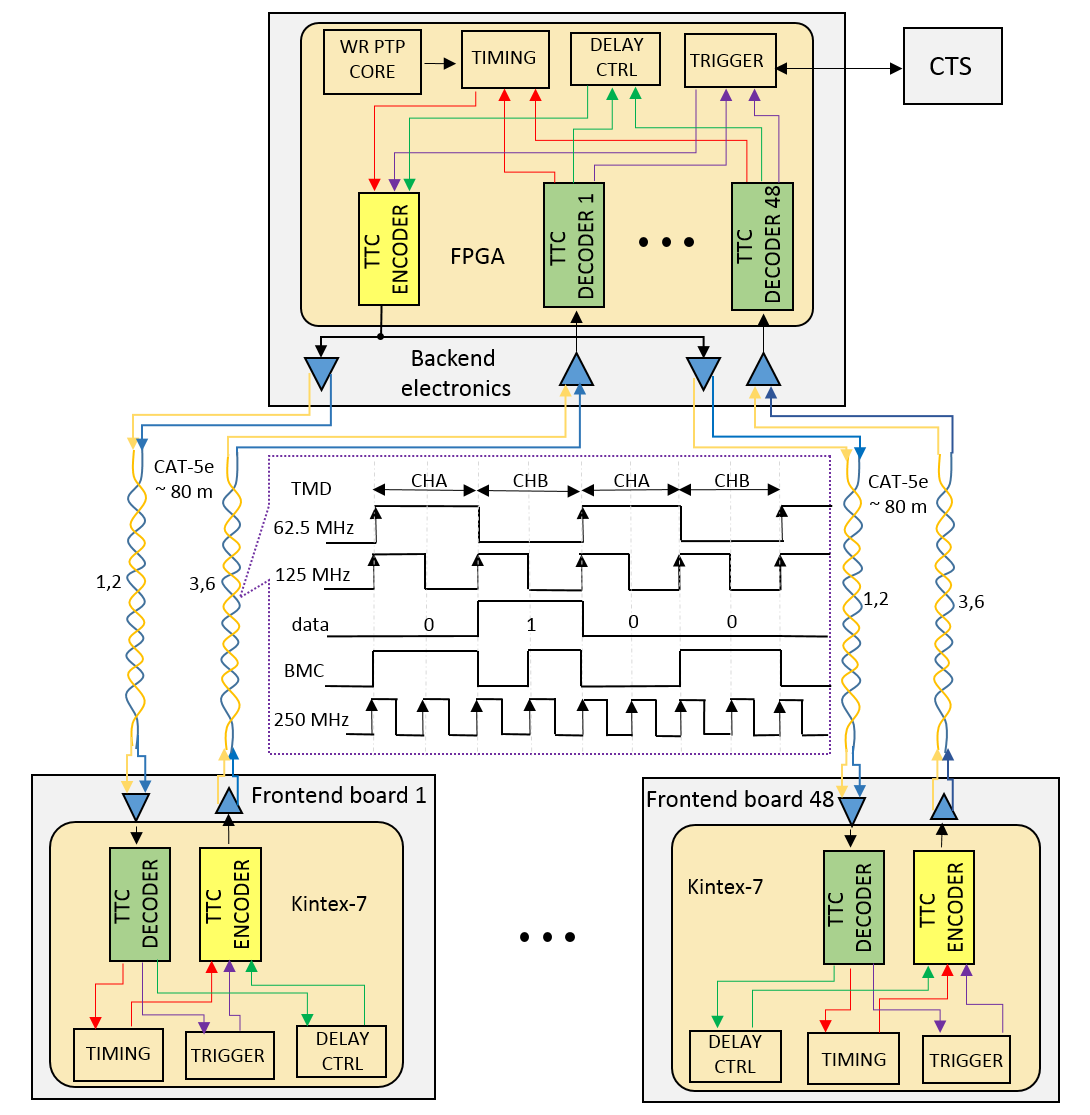}
\caption{Timing system physical and data link layers.}
\label{pedre7}
\end{figure}

PTP relies on a multicast communication model that ensures bidirectional and asynchronous messaging between master and slaves. Ethernet is the mostly used interconnection model but PTP is not limited to Ethernet. The interconnection system proposed exploits a couple of twisted pairs, in a CAT-5e cable, as a transmission medium between the master (BEC) and each of the 48 slaves connected to it (GCUs). An overview is given in Figure \ref{pedre7}. 

The physical and data link layers are based on the CERN's timing, trigger and control system concept \cite{TTC}. 
The TTC encoder and decoder implement a simple data link layer whose primary tasks are framing and error checking and correction with the Hamming codes. It does not implement data flow control and handshaking mechanisms. 
Two communication channels are time division multiplexed (TDM). Channel A is reserved for future delay calibration developments while channel B is used to encode broadcast commands consisting of 16-bit frames decoded by all receivers, and, individually addressed commands consisting of 42-bit frames. These long frames contain a header, the receiver identification number, the receiver internal address and data fields.  
At the physical layer, data is BiPhase Mark encoded (BMC) to ensure a DC balanced transmission and a self-clocking solution.

The network and transport layers, with reference to the standard Ethernet stack, are not provided by the TTC system whose aim is to implement a simple, deterministic and low latency bidirectional communication channel between BEC and GCUs. The TTC model satisfies the requirements of the trigger system, timing system, and the serial link synchronization system (delay control in the figure).
The trigger request and validation messages have the highest priority since their latency must be bounded, with the upper bounds imposed by the frontend data buffer capability. 

\subsection{Clock Syntonization}
The JUNO timing system foresees the global clock signal to be distributed to the frontend nodes as encoded information in the TTC messages. The local clock in any GCU refers to the syntonized copy of the global clock recovered from the data stream generated by the master. The syntonization is based on a clock and data recovery Integrated Circuit (IC) that guarantees that any local clock is locked in frequency with the global clock. This makes the slaves immune to medium and long term frequency drifts that manifest as a linearly increasing phase difference and a cumulative error on the local time count.  

\subsection{PTP Digital Design Overview}

\begin{figure}[!t]
\centering
\includegraphics*[width=90mm]{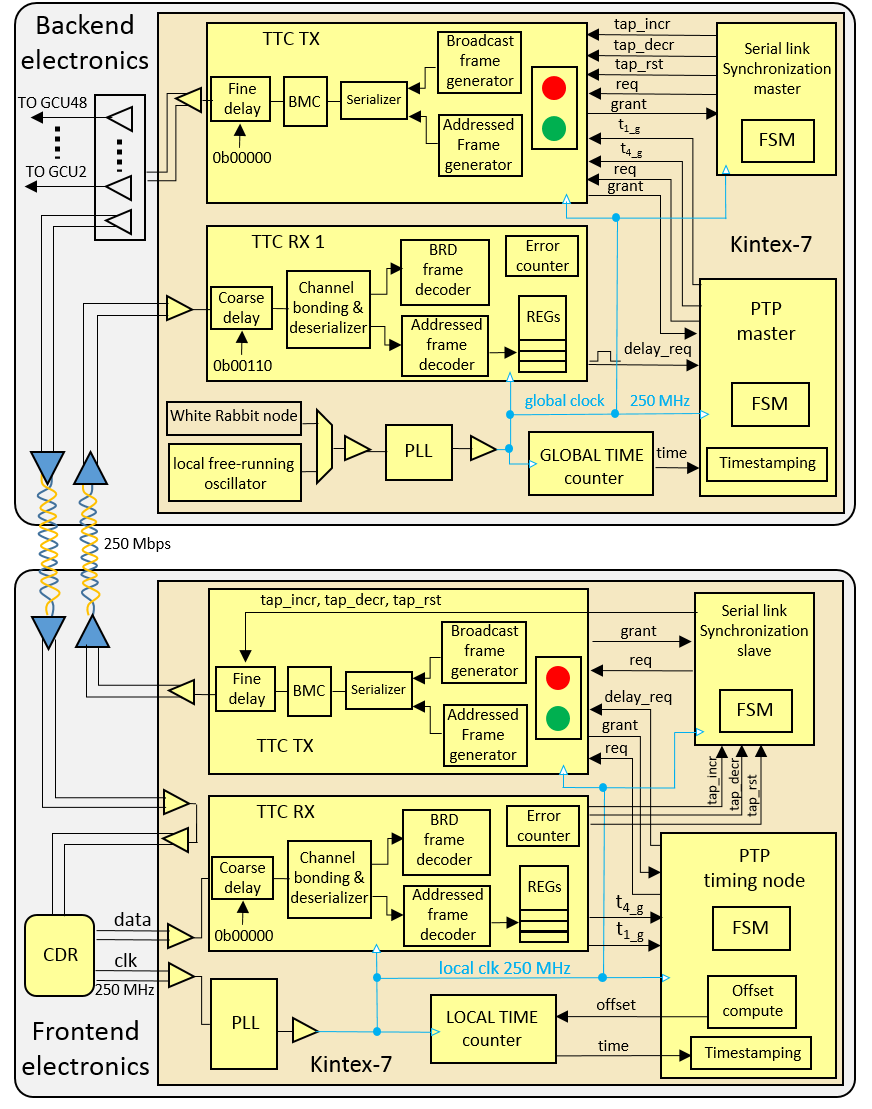}
\caption{RTL design overview.}
\label{pedre8}
\end{figure}

A complete overview of the digital circuit that implements the offset correction mechanism is given in Figure \ref{pedre8}. 
This timing system requires the availability of two FPGAs (one in the backend card and one in the frontend card) and a full duplex communication channel between the two. The choice of the communication medium bounds the maximum admissible distances between master and slave nodes. The proposed design is based on a CAT-5e unshielded twisted pair (UTP) cable, therefore, the maximum distance is about 100 m with the support of two cable driver-receiver couples. The expected data rate of 250 Mbps is well in the range of the high range (HR) general purpose I/O pin capability of the chosen FPGA, thus freeing the design of the communication physical layer from the usage of dedicated transceivers with a consequent reduction of the power consumption. The VHDL code is generic and might be synthesized for any FPGA manufacturer just replacing the I/O buffers and the clock management tiles with those provided for the family chosen. The test setup implemented exploits a Xilinx's Kintex-7 XC7K160T but the design fits comfortably in a smaller size and lower power FPGA. 
During the tests, the global clock signal has been emulated with a free running oscillator. After the integration in the WR network, it will be provided by the WR PTP core.    
Each frontend board recovers the global clock with the CDR and counts it locally. The CDR output data and clock buffers introduce a fixed latency (source of asymmetry) that has been measured using the Xilinx's ChipScope debug tool and compensated thanks to the programmable coarse delay input stage of any TTC decoder.
The offset correction protocol and messages flow have been conceived as a couple of master-slave finite state machines (FSMs).
The CERN's internal release of the TTC decoder and encoder cores have been revised and optimized to accommodate the custom timing system requirements.
The TTC has no handshaking mechanism, therefore, the PTP master and slave cores implement a watchdog that takes back the FSM to idle state in case that a message is not correctly delivered and the offset correction procedure stalls.
The PTP master scheduler follows a round-robin algorithm to address the synchronization procedure to the timing receiver nodes sequentially. The synchronization cycle is periodical.

\section{Serial Link Synchronization}

\begin{figure}[!t]
\centering
\includegraphics*[width=90mm]{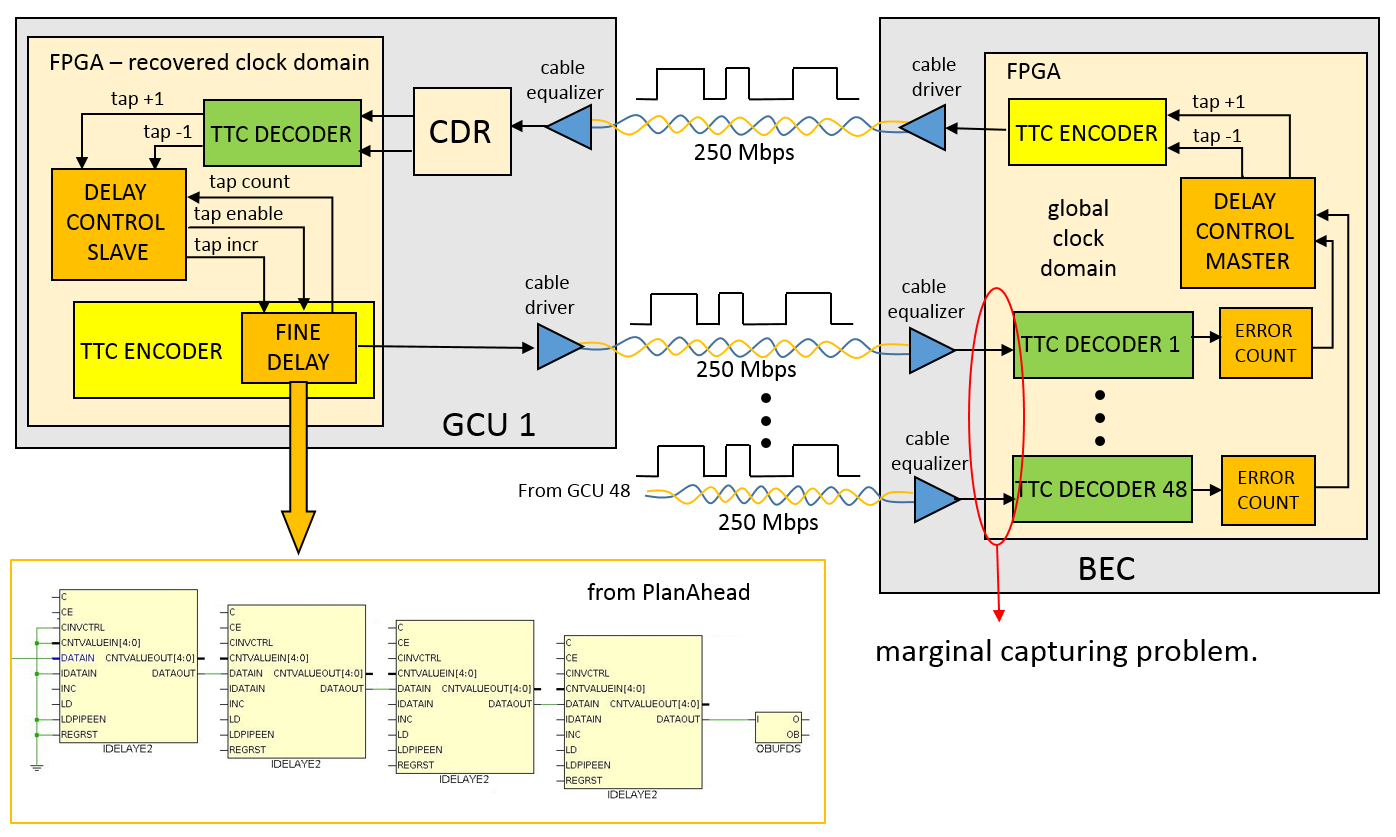}
\caption{Serial data streams synchronization}
\label{pedre9}
\end{figure}

\begin{figure}[!t]
\centering
\includegraphics*[width=86mm]{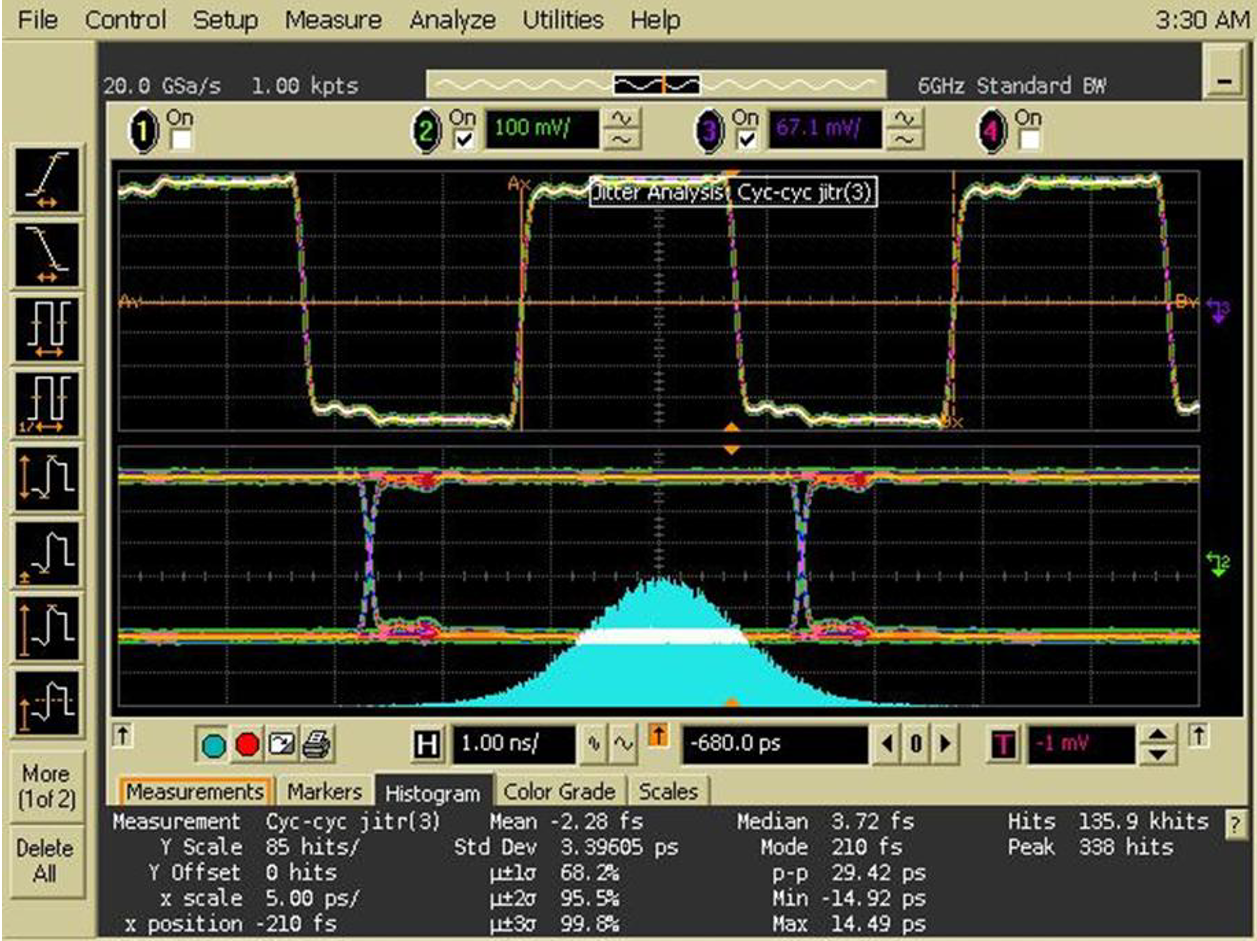}
\caption{250 MHz clock recovered on the GCU, CDR data output eye diagram and jitter measurement.}
\label{pedre10}
\end{figure}

The serial link synchronization procedure addresses the marginal capturing into the FPGAs and it is crucial to establish a reliable communication channel between backend and frontend electronics \cite{marginal capturing}. 
As shown in Figure \ref{pedre9}, in the frontend electronics the CDR chip automatically locks on the input data stream and tracks the phase of the input data in order to shift the recovered clock to the best sampling point minimizing the possibility of having the marginal capturing phenomena. 
Figure \ref{pedre10} shows the recovered clock together with its jitter histogram and the data eye diagram. As visible the setup and hold times are met and the cycle-to-cycle jitter standard deviation is 3.4 ps.

\begin{figure}[!t]
\centering
\includegraphics*[width=90mm]{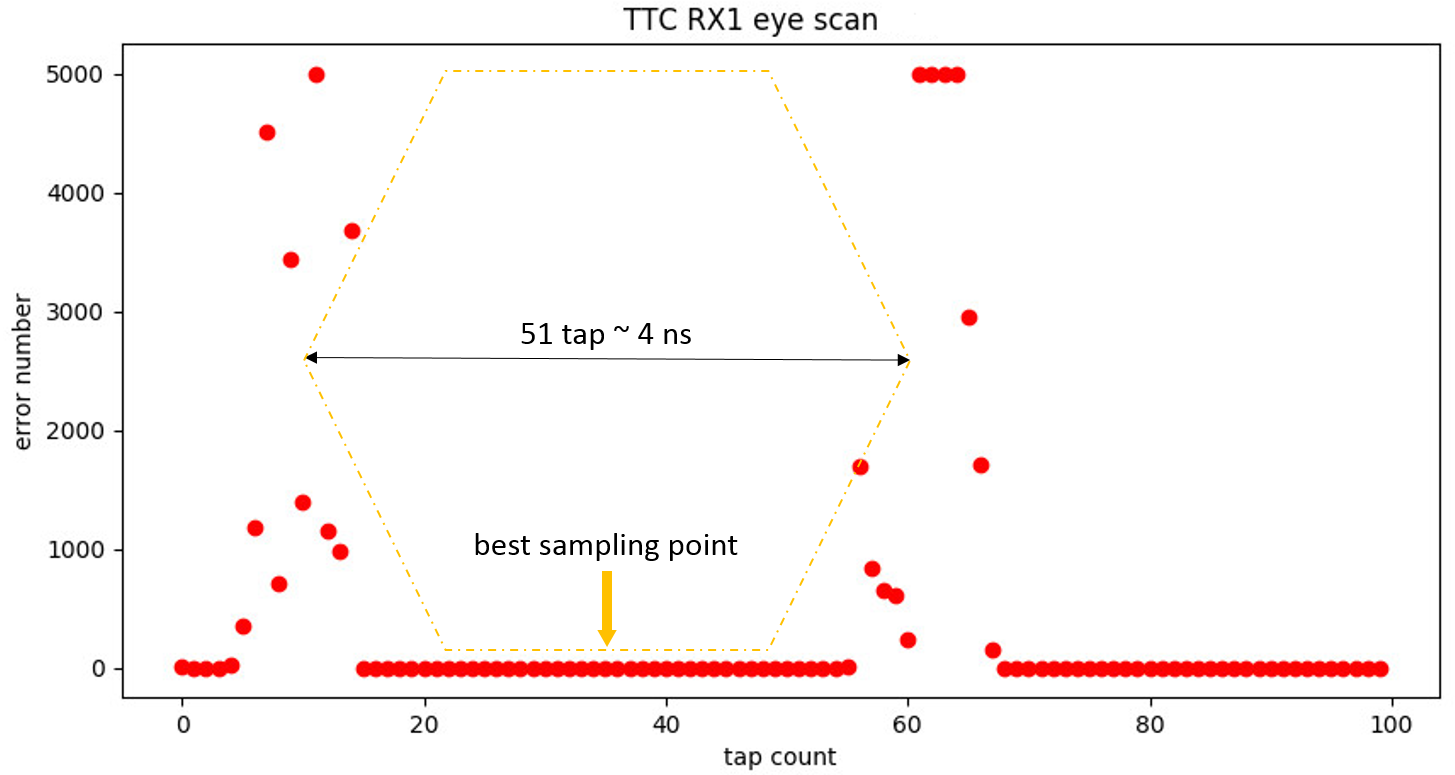}
\caption{Bathtub plot of the LVDS serial data stream capturing.}
\label{pedre11}
\end{figure}

In the backend electronics, the 250 Mbps low-voltage differential signaling (LVDS) serial data streams synchronization is more complex since 48 data streams in different phase relationship must be synchronized with the global clock domain to minimize the risk of marginal capturing that may compromise the communication stability. 
Techniques normally adopted to minimize the probability of metastability in digital designs, like synchronization registers, are not a feasible solution when the important information lies in a sequence of bits. The issue has been addressed using a cascade of 4 programmable fine delay primitives, IDELAYE2 (ODELAYE2 primitives cannot be cascaded because their output drives the corresponding I/O block and cannot be routed to the internal FPGA logic).  
Each IDELAYE2 primitive is a 31-tap wraparound selectable delay with a calibrated tap resolution of about 78 ps. 
This fine delay block is placed at the output of the TTC encoder in any frontend board, and its tap count is remotely incremented/decremented from the master calibration procedure running in the backend FPGA. The maximum delay of the chain amounts to about 9.6 ns, enough to scan two complete bit periods. The data stream input to the backend FPGA is delayed incrementally in steps of 78 ps and plotting the TTC frame error count versus the tap count one can get the information about the eye opening and the best sampling point as illustrated in Figure \ref{pedre11}.
Running at 250 Mbps the expected data eye width is 4 ns that correspond to about 51 taps.
The clock synchronization procedure cannot start until this
calibration is completed and the channel is error free.
Establishing a reliable bidirectional communication between the master and all the timing receiver nodes is essential in a 18000 channels setup.

\section{Clock Synchronization Procedure and Results Achieved}
 
The clock synchronization solution proposed has been tested and fully characterized in a test setup composed by one master and three timing receiver nodes. In sequence, these are the main steps to get an accurate copy of the global time at frontend level:
\begin{itemize}
\item power up the BEC and GCUs boards. The power up sequence of the boards is not a concern.
\item The BEC card locks with the global clock signal and starts to count the global time and broadcasts periodical \textit{idle} commands.
\item Each GCU locks to the recovered global clock copy and starts counting the time locally. 
\item Each GCU starts the channel identification procedure necessary to decode the TTC commands. As soon as an \textit{idle} command is correctly decoded, the channel aligned flag is set to '1'. 
\item The BEC broadcasts to all GCUs an error reset command. Upon having decoded the error reset command, all the TTC decoders into the frontend boards are expected to be error free. 
\item Enable the serial link synchronization procedure.
\item Check that all the communication channels are error free. 
\item Enable the clock synchronization procedure in the master and in all GCUs. 
\item Once enabled, the offset correction mechanism is periodical.
\end{itemize}

\begin{figure}[!t]
\centering
\includegraphics*[width=90mm]{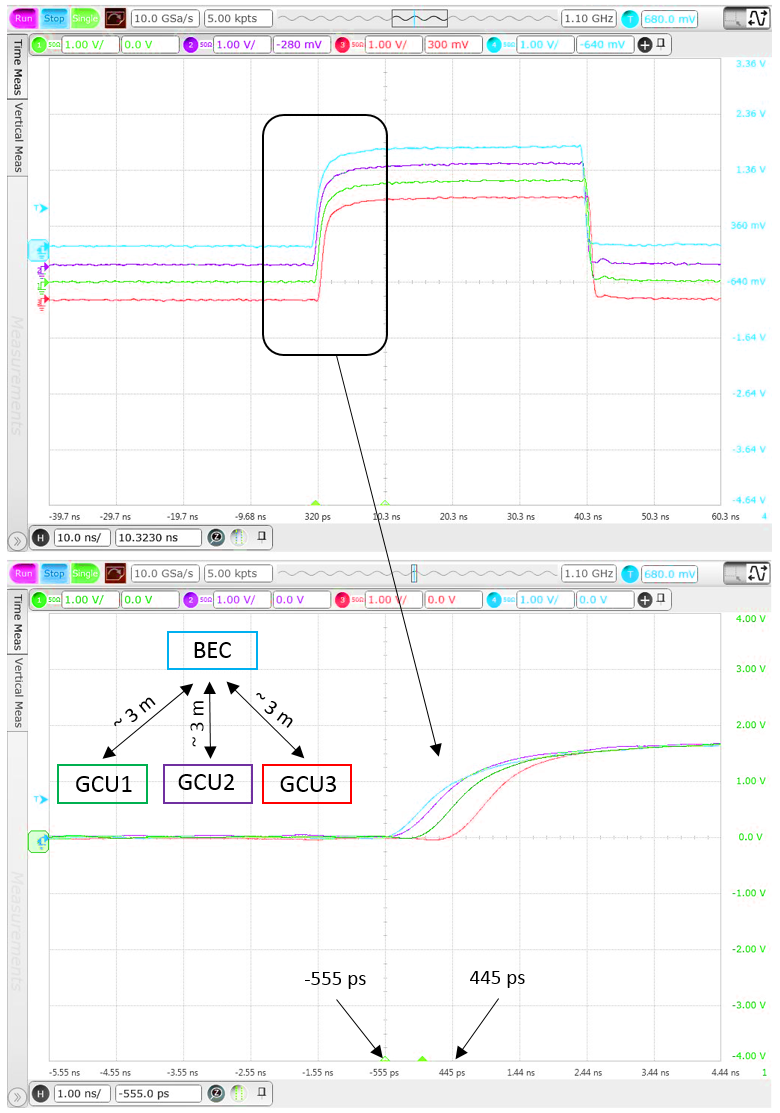}
\caption{Time accuracy achieved with about 3 m of CAT-5e cable.}
\label{pedre12}
\end{figure}

The outcome of the offset correction procedure has been verified using an oscilloscope as an external observer. The backend and frontend boards have been programmed to generate a pulse at a scheduled time, and, the output pulses observed with the oscilloscope are shown in Figure \ref{pedre12}.
The pulses are aligned within 1 ns. As expected there is no control on the phase relation between the global clock signal and the local clock signals. 

\begin{figure}[!t]
\centering
\includegraphics*[width=90mm]{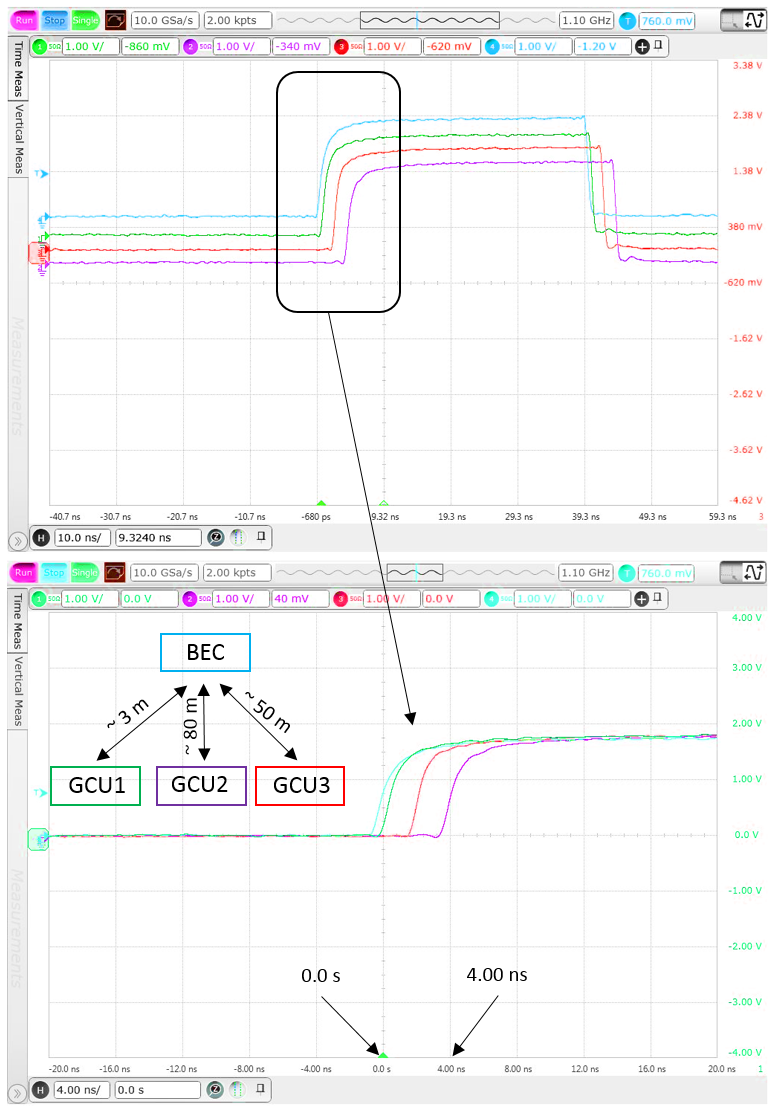}
\caption{Time accuracy achieved with a 3 m long cable channel 1, 80 m long cable channel 2, 50 m long cable channel 3.}
\label{pedre13}
\end{figure}

\begin{figure}[!t]
\centering
\includegraphics*[width=88mm]{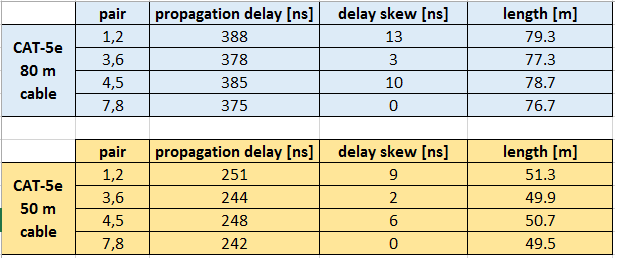}
\caption{CAT-5e UTP cables analysis.}
\label{pedre14}
\end{figure}

The test has been repeated with three cables of different length to reproduce a condition similar to the final installation on the field. The result is shown in Figure \ref{pedre13}.
The time accuracy achieved is well within the requirements of $\mathit{\pm}$ 16 ns, but, the time offset of the GCU2 is slightly larger than the 250 MHz recovered clock period. The offset error is induced by the asymmetry \cite{asymmetry 1}. The only source of asymmetry in the design (not compensated) is the CAT-5e copper cable. Typical propagation delay for CAT-5e UTP is in the order of few ns per meter and the standard specifies that a 100 m cable might have a delay skew between pairs up to 50 ns, due to the different twist rate. Figure \ref{pedre14} displays the cable analysis performed on the 80 m and the 50 m cables used to test the timing system.
The asymmetry between the pairs 1,2 and 3,6 of the 80 m cable is 10 ns.  This asymmetry generates the clock period offset error observed.
Without any compensation mechanism, the offset error introduced by the cable asymmetry may be up to 25 ns. In this worst case scenario, the local clock would be out of specification.

\subsection{Accuracy Improvements}

The compensation of the asymmetrical latency introduced by the cable is necessary to claim a time resolution of $\mathit{\pm}$ 4 ns using copper cables.
 
In a context like JUNO where cable layout cannot be changed after installation, asymmetry could be measured at the cable supplier premises and then manually compensated using coarse and fine delay primitives included in the firmware.

Dedicated hardware support in frontend and backend cards could provide the possibility of swapping transmit and receive paths, hence allowing an automatic measurement of cable length imbalance and consequent compensation \cite{asymmetry 2}.

Where applicable, the digital implementation of PTP will benefit a lot from a TTC optical distribution. The communication medium asymmetry would then be negligible obtaining a $\mathit{\pm}$ 4 ns timing system over an extended transmission range. 

\section{Conclusion}
\balance
A fully hardware implementation of PTP for offset measurement and compensation has been developed and tested. The digital design proposed includes the TTC system as physical and data link layer to realize a fully duplex and deterministic latency communication channel between master and slaves and enables the synchronization of thousands of timing receiver nodes with a precision of $\mathit{\pm}$ one clock period. The test setup described in the paper is based exclusively on the presence of the FPGA technology on the backend and on the frontend electronics, with a communication medium consisting of a standard CAT-5e UTP cable. 
The results achieved confirm that the implemented timing system is a cost effective solution to extend the time accuracy to $\mathit{\pm}$ 4 ns without complex calibration procedures. 
The fully hardware design together with the deterministic multicast communication system gets rid of the asymmetries introduced by classical PTP software implementations over standard Ethernet networks with the consequent performance improvements. 
The only source of asymmetry of the proposed timing system is the physical medium that, if not compensated, may cause offset errors. 

%\ifCLASSOPTIONcaptionsoff
%  \newpage
%\fi
%\IEEEtriggeratref{15}

%\vspace{0.5cm}



\end{document}